\documentclass[a4paper,11pt]{article}

\usepackage{jheppub} 

\usepackage[T1]{fontenc} 

\def\nn{\nonumber\\ }
\def\rd{{\rm d}}
\def\abs#1{\left| #1 \right| }
\def\hsix{ \mathcal{H}^{(6)}}
\def\b#1{b_0^{(#1)}}

\title{Renormalization Group Scaling of Higgs Operators and $h \to \gamma \gamma$ Decay}

\author[a,b]{Christophe Grojean,}

\author[c]{Elizabeth E.~Jenkins,}

\author[c]{Aneesh V.~Manohar,}

\author[a,1]{Michael Trott}\note{Corresponding author.}

\affiliation[a]{Theory Division, Physics Department, CERN, CH-1211 Geneva 23, Switzerland}
\affiliation[b]{ICREA at IFAE, Universitat Aut{\`o}noma de Barcelona, 08193 Bellaterra, Barcelona, Spain}
\affiliation[c]{Department of Physics, University of California at San Diego, 9500 Gilman Drive,\\ La Jolla, CA 92093-0319, USA}

\emailAdd{christophe.grojean@cern.ch}
\emailAdd{ejenkins@ucsd.edu}
\emailAdd{amanohar@ucsd.edu}
\emailAdd{michael.trott@cern.ch }

\abstract{
We compute the renormalization of dimension six Higgs-gauge boson operators that can modify the $h \to \gamma \gamma$ rate at tree-level. Operator mixing is shown to lead to an important modification of new physics effects which has been neglected in past calculations. We also find that the usual formula for the $S$ oblique parameter contribution of these Higgs-gauge boson operators needs additional terms  to be consistent with renormalization group evolution. We study the implications of our results for Higgs phenomenology and for new physics models which attempt to explain a deviation in the $h \to \gamma \gamma$ rate. We derive a new relation between the $S$ parameter and the $h \to \gamma\gamma$ and $h \to \gamma Z$ decay rates.
}

\begin{document} 
\maketitle
\flushbottom

\section{Introduction}

The discovery of a new boson~\cite{:2012gk,:2012gu} with a mass around 126~GeV, based on excess events in several Higgs search channels at the LHC, has reshaped the field of particle physics. The leading candidate, by far, for the observed boson is the standard model (SM) Higgs boson. It is important to study the production and decay rates of this new particle with high precision to verify that they agree with the predictions of the SM: computing the Higgs production and decay rates with higher precision within the SM, and performing precise computations in continuous deformations away from the SM are necessary.
At stake in these studies are naturalness arguments that have been widely used to predict that there should be new physics (NP) associated with electroweak (EW) symmetry breaking at scales not far above the EW scale. 

One aspect of the properties of the observed resonance that has attracted some attention is the apparent excess in the  $\Gamma(h \rightarrow \gamma  \gamma)$ decay rate. This excess may be just a statistical fluctuation, or it may be due to the effects of NP modifying the properties of a SM Higgs. Although this deviation from the SM expectation has received the most attention to date, the properties of the observed resonance are not known experimentally to be in detailed agreement with SM expectations in many search channels. If deviations of the properties of the observed state from SM expectations become statistically significant in the signal strengths for the decays $h \rightarrow \gamma  \gamma, WW, ZZ, Z  \gamma$, a program of precision Higgs phenomenology will be key to unraveling the physics beyond the SM.

In this paper, we will assume that the new boson corresponds to the Higgs boson and that the NP scale is at least a few hundred GeV, so that the effect of new physics can be captured by adding higher-dimension operators to the SM Lagrangian.  If NP influences the properties of the observed boson, one must consistently calculate the relationship between the Wilson coefficients of the higher dimensional operators, at the low-energy EW scale $\sim v$ and the high-energy scale $\Lambda$ ---
which corresponds to the mass scales of the NP states that are integrated out of the effective theory. Systematically relating the
Wilson coefficients at these different scales requires determining the anomalous dimensions of the operator basis,
including the effects of operator mixing. In this paper, we determine the anomalous dimension matrix for a set of 
operators that affect the decay of the SM Higgs to $W  W, ZZ, Z  \gamma$ and $\gamma  \gamma$.
The operator basis we focus on leads to tree-level modifications of the $\gamma  \gamma$ and $Z  \gamma$ Higgs decays, which first arise at one loop in the SM, and it also is constrained by electroweak precision data (EWPD). We show that earlier investigations \cite{Grinstein:1991cd, Hagiwara:1993ck, Hagiwara:1993qt, Alam:1997nk, Han:2004az} relating the $S$ parameter to higher dimensional operators correctly capture some of the scale dependence of the operators. However, these results need to be modified to take into account the full scale dependence of the operators determined by the renormalization group equations (RGE). We study the constraints on operator mixing from the $S$ parameter in detail, deriving a new relation between the Higgs decay rates and the $S$ parameter.

The operator mixing matrix computed here allows for the identification of a new mechanism by which NP contributes to $h \to \gamma  \gamma$  and $h \to Z \gamma$ decays. 
These new contributions of NP to one-loop Higgs decays can be much larger than naively expected when considering a RGE effect --- as we show in an explicit example. The key point is that an operator that is matched onto at tree level when integrating out a NP sector, that subsequently mixes with the operator corresponding to the one-loop Higgs production or decay process can lead to a NP contribution that is of the same order as a direct matching contribution. Our results demonstrate this general point: {\it systematically accounting for the scale dependence of the NP induced operators is essential for correctly calculating a one-loop Higgs process in an effective action that reproduces the infrared of a NP theory extension of the SM.}

The outline of this paper is as follows. Section~\ref{frameworks}  sets up our notation and defines the operator basis that we renormalize.
In Section~\ref{sec:anom}, we give the anomalous dimension matrix of the dimension-six Higgs-gauge boson operators. The implications of our results for LHC phenomenology, and for electroweak precision constraints are given in  Section~\ref{pheno}.  Finally, we give our conclusions in Section~\ref{concl}.

\section{The Operator Basis}\label{frameworks}

We assume that at the scale of the Higgs mass, $\mu \sim M_h \sim 126$~GeV, the theory is represented by the $SU(3) \times SU(2) \times U(1)$ standard model (SM) with the minimal Higgs sector. The new physics effects are given by gauge invariant local operators in terms of the SM fields. The lowest dimension operators are dimension-five lepton-number violating operators which give rise to neutrino masses. The operators which first affect the properties of the Higgs boson occur at dimension six. A complete classification of the dimension-six operators in the standard model is given in Refs.~\cite{Buchmuller:1985jz,Grzadkowski:2010es}, the latter of which finds that there are 59 independent operators (assuming baryon number conservation) after eliminating redundant operators using the equations of motion. The choice of independent operators is not unique, since certain linear combinations vanish by the equations of motion, and are thus effectively of higher dimension.

In this paper, we will consider the impact of the following dimension-six operators modifying the standard model Hamiltonian,
\begin{align}
\hsix = -\mathcal{L}^{(6)} &= c_{G} \, \mathcal{O}_{G}+c_{B} \, \mathcal{O}_{B} + c_{W} \, \mathcal{O}_{W} + c_{W\!B} \, \mathcal{O}_{WB} \nn
& + \widetilde{c}_{G} \, \widetilde{\mathcal{O}}_{G} + \widetilde{c}_{B} \, \widetilde{\mathcal{O}}_{B} + \widetilde{c}_{W} \, \widetilde{\mathcal{O}}_{W} + \widetilde{c}_{WB} \, \widetilde{\mathcal{O}}_{WB}\,.
\label{h6}
\end{align}
The Hamiltonian $\hsix$ is generated by new physics at some scale $\Lambda$. The
operator basis for $\hsix$  is (using the notation of Ref.~\cite{Manohar:2006gz,Manohar:2006ga})
\begin{equation}
\begin{aligned}
\mathcal{O}_{G} &=  \frac{g_3^2}{2 \, \Lambda^2} \, H^\dagger \,  H \, G_{\mu\, \nu}^A G^{A\, \mu \, \nu}, & \hspace{1cm}
\widetilde{\mathcal{O}}_{G} &=  \frac{g_3^2}{2 \, \Lambda^2} \, H^\dagger \,  H \,  G_{\mu\, \nu}^A \widetilde G^{A\, \mu \, \nu}, \\
\mathcal{O}_{B} &=  \frac{g_1^2}{2 \, \Lambda^2} \, H^\dagger \,  H \, B_{\mu\, \nu} B^{\mu \, \nu}, & 
\widetilde{\mathcal{O}}_{B} &=  \frac{g_1^2}{2 \, \Lambda^2} \, H^\dagger \,  H \, {B}_{\mu\, \nu} \widetilde B^{\mu \, \nu}, \\
\mathcal{O}_{W} &=  \frac{g_2^2}{2 \, \Lambda^2} \, H^\dagger \,  H \, W^a_{\mu\, \nu} W^{a\,\mu \, \nu}, &
\widetilde{\mathcal{O}}_{W} &=  \frac{g_2^2}{2 \, \Lambda^2} \, H^\dagger \,  H \, {W}^a_{\mu\, \nu}  \widetilde W^{a\,\mu \, \nu}, \\
\mathcal{O}_{WB} &=  \frac{g_1 \, g_2}{2 \, \Lambda^2} \, H^\dagger \, \sigma^a \, H \, W^a_{\mu \, \nu} B^{\mu\, \nu},  &
\widetilde{\mathcal{O}}_{WB} &=  \frac{g_1 \, g_2}{2 \, \Lambda^2} \, H^\dagger \, \sigma^a \, H \, W^a_{\mu \, \nu} \widetilde B^{\mu\, \nu}  .
\end{aligned}
\label{ops}
\end{equation}
Here, $g_1$, $g_2$ and $g_3$ are the standard model gauge couplings,  $B_{\mu\nu}$, $W^a_{\mu \nu}$ and $G^A_{\mu \nu}$ are the corresponding field-strength tensors,
and $\sigma^a$ are the Pauli matrices for weak isospin. The operators $\mathcal{O}_i$ are $CP$-even, and $\widetilde{\mathcal{O}}_i$ are $CP$-odd. The dual field-strength tensors are defined by $\widetilde{F}_{\mu \nu} = (1/2) \, \epsilon_{\mu \nu \alpha \beta} F^{\alpha \beta}$, for $F=B,W^a,G^A$. Note that the $\ {}\widetilde{}\ $ can be on either field-strength, since
$F_{1\, \mu \nu} \widetilde F_{2\, \mu \nu} = \widetilde F_{1\, \mu \nu}  F_{2\, \mu \nu}$.  This observation will be useful later. Only the product $c_i/\Lambda^2$ enters $\hsix$, but it is useful to write the operators in the form of Eq.~(\ref{ops}) so that the coefficients $c_i$ in $\hsix$ are dimensionless. A naive dimensional estimate~\cite{Manohar:1983md} gives $c_i$ of order unity. Nevertheless the relative importance of the various operators will depend on the power counting of the NP model considered --- we will discuss this point in more detail in 
Section~\ref{NewNP}.

The Higgs doublet field $H$ has hypercharge $Y=+1/2$, and the Higgs potential is
\begin{align}
V &= \lambda \left(H^\dagger \, H - \frac{v^2}{2} \right)^2\,.
\end{align}
With this normalization convention, $v \sim 246$~GeV and $M_h^2=2\lambda v^2$. Yukawa couplings are normalized in the usual way, so that the fermion masses $m_f$ are given in terms of the Yukawa couplings $y_f$ by $m_f =y_f v/\sqrt 2$.

In Section~\ref{sec:anom}, we compute the anomalous dimension matrix for the subset of dimension-six operators in Eq.~(\ref{ops}). 
The operator basis Eq.~(\ref{ops}) is closed under renormalization at one loop (for the diagrams in Fig.~\ref{fig:graphs}).
The reason we focus upon the operators in $\hsix$ is that they contribute at tree level to  $\gamma \gamma$ and $Z \gamma$ Higgs decays, which are one-loop processes in the standard model. Thus, these operators are particularly important for the present analysis and current phenomenology. 

We note that other dimension-six operators, such as
\begin{align}\label{2.4}
\mathcal{O}_{DB} &=  (D^\mu H)^\dagger \,  (D^\nu H) \, g_1 \, B_{\mu \, \nu},   &  \mathcal{O}_{DW} &= (D^\mu H)^\dagger \, \sigma^a \, (D^\nu H) \, g_2 \, W^a_{\mu \, \nu}, \nn
\mathcal{O}_\phi &=  \abs{H^\dagger \, D^\mu H}^2,  & {\mathcal{O}}_{WWW} &=  g_2^3 \, \epsilon_{abc} W^a_{\mu \, \nu} \, W^b_{\nu \, \rho} \, W^c_{\rho \, \mu},
\end{align}
are also of interest for precision EW phenomenology.  These neglected dimension-six operators also can mix with the operators
Eq.~(\ref{ops}) under renormalization group scaling, so a renormalization group analysis of the complete dimension-six operator basis is needed to obtain all effects.  The calculation of the 59 $\times$ 59 anomalous dimension mixing matrix of dimension-six operators is beyond the scope of the present work, but merits future investigation.   

The basis we use, given by Eq.~(\ref{ops}), is sufficient to demonstrate the point we wish to make
on RGE effects due to operators that can come about due to NP at tree level. It is well known that tree level NP effects 
can lead to contributions to the $S$ parameter, which  corresponds to $\mathcal{O}_{WB}$.
Note that while the operators ${\cal O}_{DB}$ and ${\cal O}_{DW}$ do appear in the operator basis of Refs.~\cite{Hagiwara:1993ck,Hagiwara:1993qt,Alam:1997nk} and in the strongly interacting light Higgs (SILH) basis of Ref.~\cite{Giudice:2007fh}, they do not appear in the basis of Ref.~\cite{Grzadkowski:2010es} where they have been replaced in favour of the two operators with fermionic currents:
\begin{equation}
i \left( H^\dagger \sigma^a {\overleftrightarrow { D^\mu}} H \right) \left( {\bar q}_L \gamma^\mu \sigma^a q_L + {\bar l}_L \gamma^\mu \sigma^a l_L \right), \hspace{2cm}
i \sum_{\psi} \left( H^\dagger {\overleftrightarrow { D^\mu}} H \right) \left( y_i {\bar \psi} \gamma^\mu \psi \right),
\end{equation}
where $\psi=q_L,d_R,u_R,l_L,e_R$ and $y_i$ is their hypercharge. See Ref.~\cite{Grojean:2006nn} for further discussion. These two fermionic current operators correspond to oblique corrections. It is therefore preferable to choose an operator basis
which replaces them by purely bosonic operators as in Refs.~\cite{Hagiwara:1993ck,Hagiwara:1993qt,Alam:1997nk} and in Ref.~~\cite{Giudice:2007fh}.  In the basis of Ref.~\cite{Grzadkowski:2010es}, the three operators ${\cal O}_B, {\cal O}_{W}$ and ${\cal O}_{W\!B}$ will be generated at the loop-level only. Conversely, in the basis we use, the operator $\mathcal{O}_{WB}$ can receive a tree-level matching due to NP. Thus, we consider the calculation we have performed to be sufficient to demonstrate the importance of the RGE improvement of Higgs production and decay operators when NP can lead to tree-level matching. Although the tree-level operator we demonstrate this point with, $\mathcal{O}_{WB}$, is directly bounded by EWPD to be smaller than its naive dimensional estimate,
we emphasize that in integrating out a realistic new physics sector one expects a number of tree level effects, unless the new sector is protected by an exact discrete symmetry ---
such as in an exact R parity conserving SUSY model. 

The classification of tree-level NP effects in the dimension-six operator basis was first performed in Refs.~\cite{Arzt:1994gp},
which finds 45 operators can be induced at tree level in their chosen basis. In the classification of  Ref.~\cite{Grzadkowski:2010es}, 14 (+ 25 four fermion)  operators can be induced by tree-level NP effects (when baryon number is assumed conserved).\footnote{Our chosen basis is particularly useful to make the RGE effect we are demonstrating clear. A basis-independent argument requires computing the full 59 x 59 anomalous dimension matrix of a complete operator basis.}

\section{Anomalous Dimensions}\label{sec:anom}

In this section, we compute the one-loop anomalous dimension of the new physics Hamiltonian $\hsix$. The computations are performed in the unbroken gauge theory with six dynamical quark flavours. We use  background field gauge and the $\rm \overline{MS}$ subtraction scheme in $d = 4 - 2 \, \epsilon$ dimensions. In  background field gauge, the product $g A_\mu$ is not renormalized due to background field gauge invariance, so that $Z_g Z_A^{1/2}=1$.\footnote{A review of the background field method can be found in Ref.~\cite{Abbott:1981ke}. $Z_g$ and $Z_A$ are the renormalization factors of the gauge coupling and field, $g^{(0)}=Z_g g$, $A_\mu^{(0)}=Z_A^{1/2}A _\mu$, respectively,  where the superscript ${}^{(0)}$ denotes bare quantities. For a recent review of NLO effective lagrangians see Ref.~\cite{Passarino:2012cb}.}

In a gauge theory, the operators
\begin{align}
\mathcal{O}_+&=\frac{\beta(g)}{2g} F^A_{\mu \nu} F^{A\,\mu \nu},  & \mathcal{O}_-&=g^2 F^A_{\mu \nu} \widetilde F^{A\,\mu \nu},
\label{3.1}
\end{align}
are not multiplicatively renormalized to all orders in perturbation theory. The $CP$-even operator $\mathcal{O}_+$ is not renormalized, because it is the trace of the conserved energy momentum tensor. (A review can be found in Ref.~\cite{Adler:1982ri}.) The $CP$-odd operator $\mathcal{O}_-$ is not multiplicatively renormalized, because it is multiplied by the $\theta$-angle in the Lagrangian, and $\theta$ is periodic with periodicity $2\pi$.\footnote{It can mix with the divergence of the axial current. See Appendix C of Ref.~\cite{Kaplan:1988ku}.} 

In the standard model, we have multiple gauge fields, so the theorem on $\mathcal{O}_+$  only applies to the sum of the three gauge contributions,
\begin{align}
\mathcal{O}_+&=\frac{\beta_1(g_1)}{2g_1} B_{\mu \nu} B^{\mu \nu} + \frac{\beta_2(g_2)}{2g_2} W^a_{\mu \nu} W^{a\,\mu \nu}+ \frac{\beta_3(g_3)}{2g_3} G^A_{\mu \nu} G^{A\,\mu \nu}.
\label{3.2}
\end{align}
The coupling between the different gauge operators in Eq.~(\ref{3.2}) only occurs through fermion or scalar loops, so at the one-loop level, the separate terms are not renormalized. Thus, at one-loop, we can use the result that both $g^2 F^A_{\mu \nu} F^{A\,\mu \nu}$ and $g^2  F^A_{\mu \nu} \widetilde F^{A \, \mu \nu}
$ are not renormalized, which provides a useful check of our computation.

\begin{figure}
\begin{center}
\begin{eqnarray*}
\begin{array}{cccc}
&\hspace{0.2cm} \includegraphics[height=3cm]{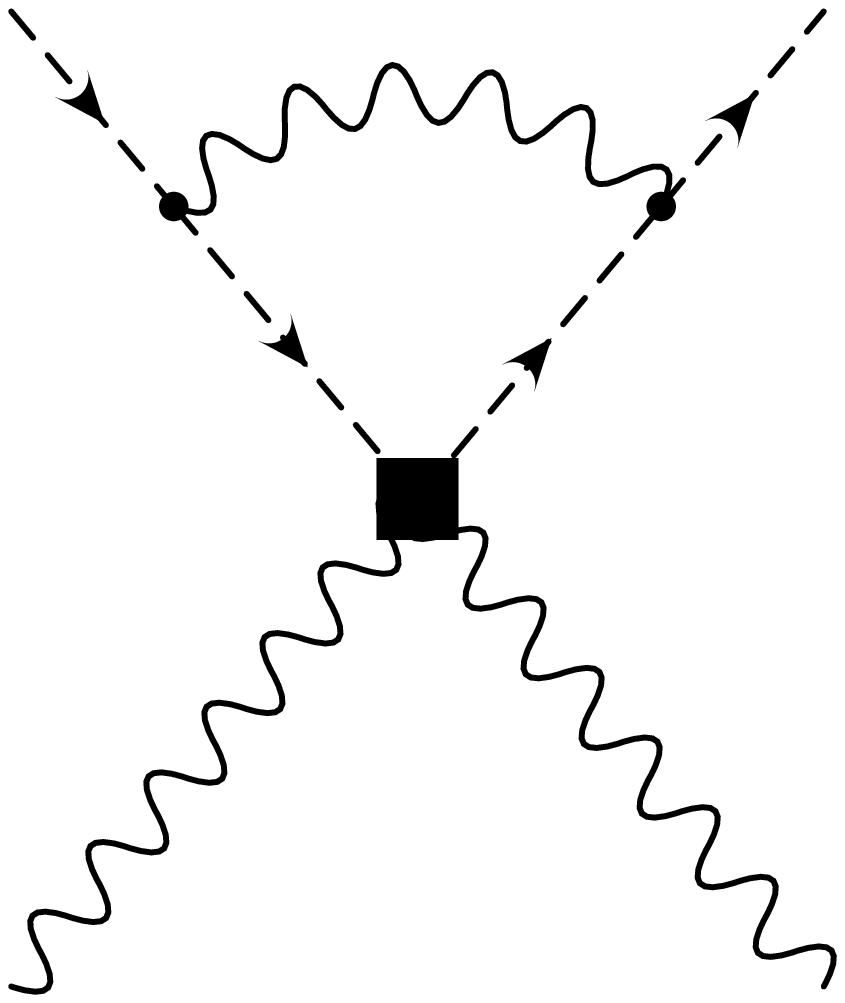} \hspace{0.2cm} &\hspace{0.2cm}  \includegraphics[height=3cm]{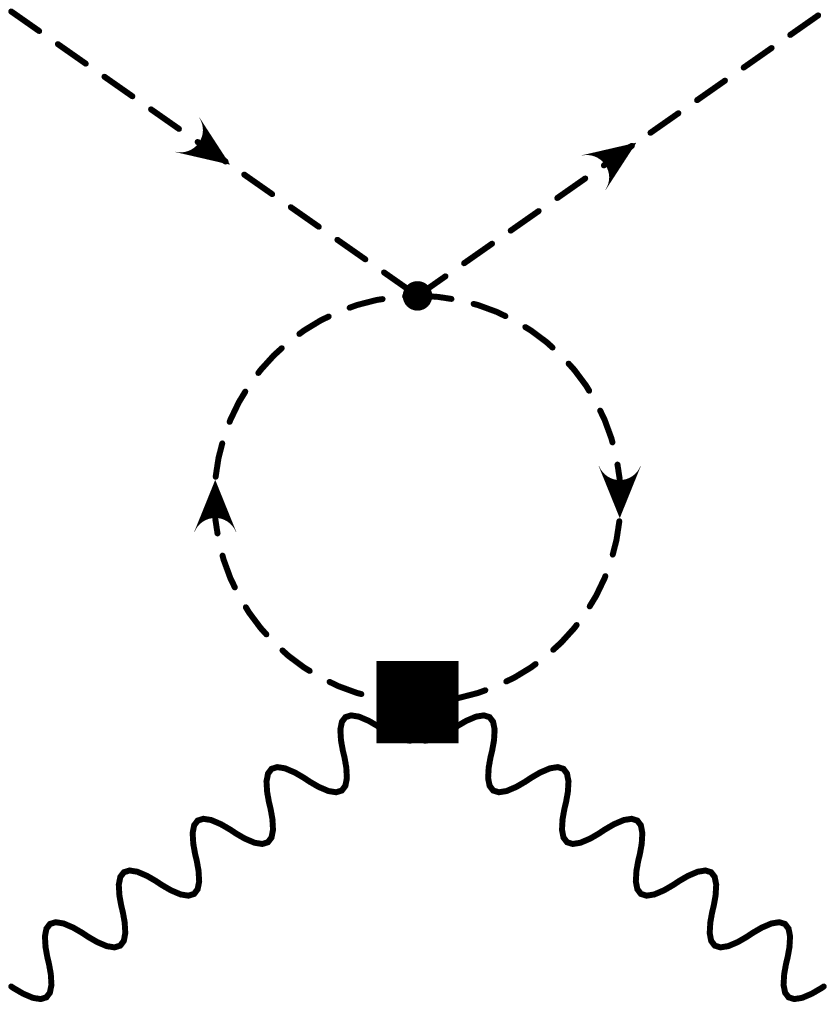}\hspace{0.2cm} \\
& (a) & (b) \\[20pt]
\hspace{0.2cm} \includegraphics[height=3cm]{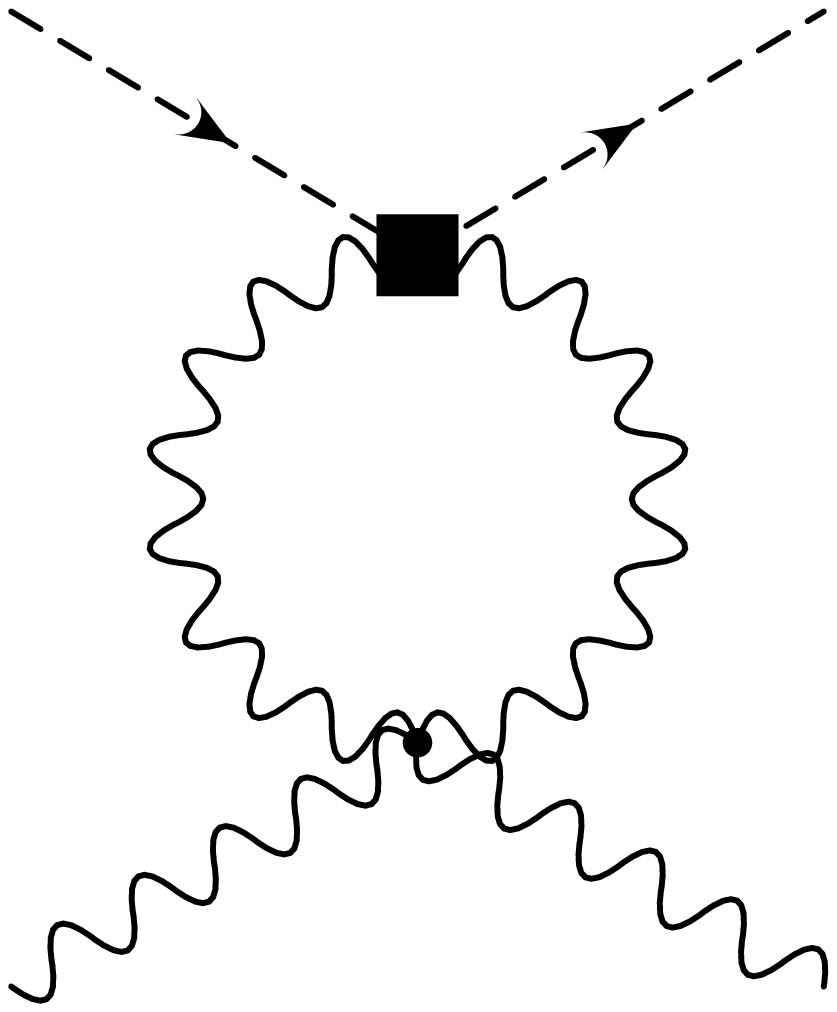} \hspace{0.2cm} &\hspace{0.2cm}  \includegraphics[height=3cm]{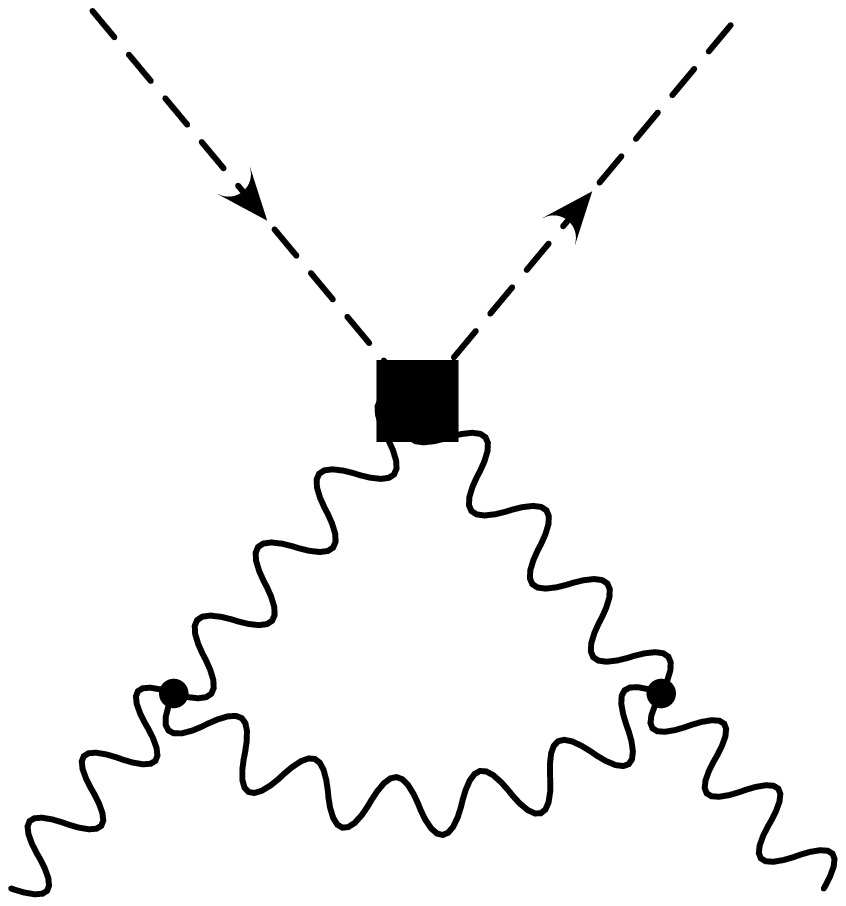} \hspace{0.2cm} &\hspace{0.2cm}
\includegraphics[height=3cm]{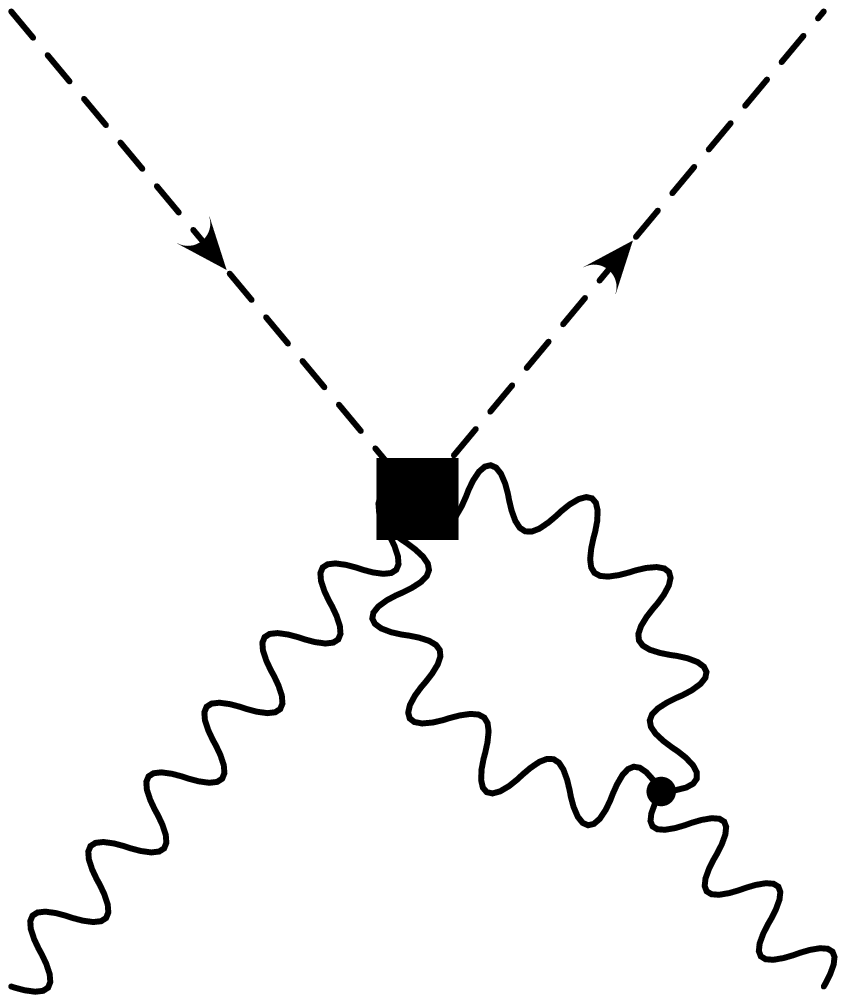} \hspace{0.2cm} & \hspace{0.2cm} \includegraphics[height=3cm]{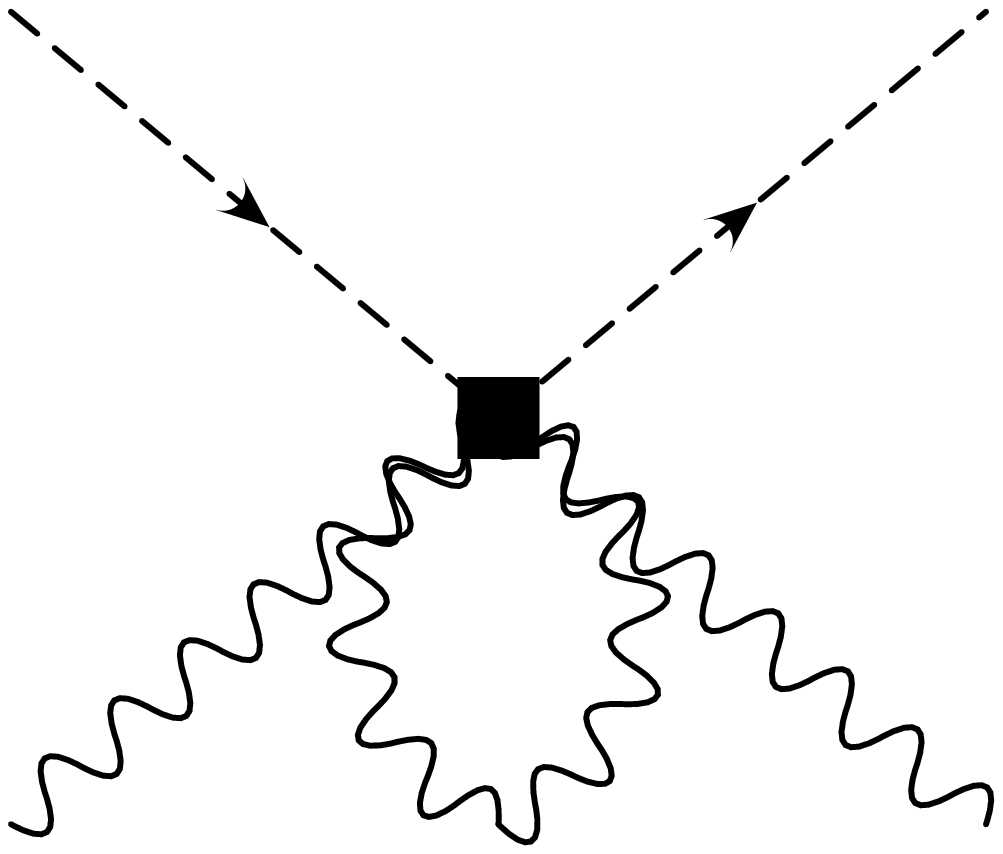} \hspace{0.2cm} \\
(c) & (d) & (e) & (f) \\[20pt]
\hspace{0.2cm} \includegraphics[height=3cm]{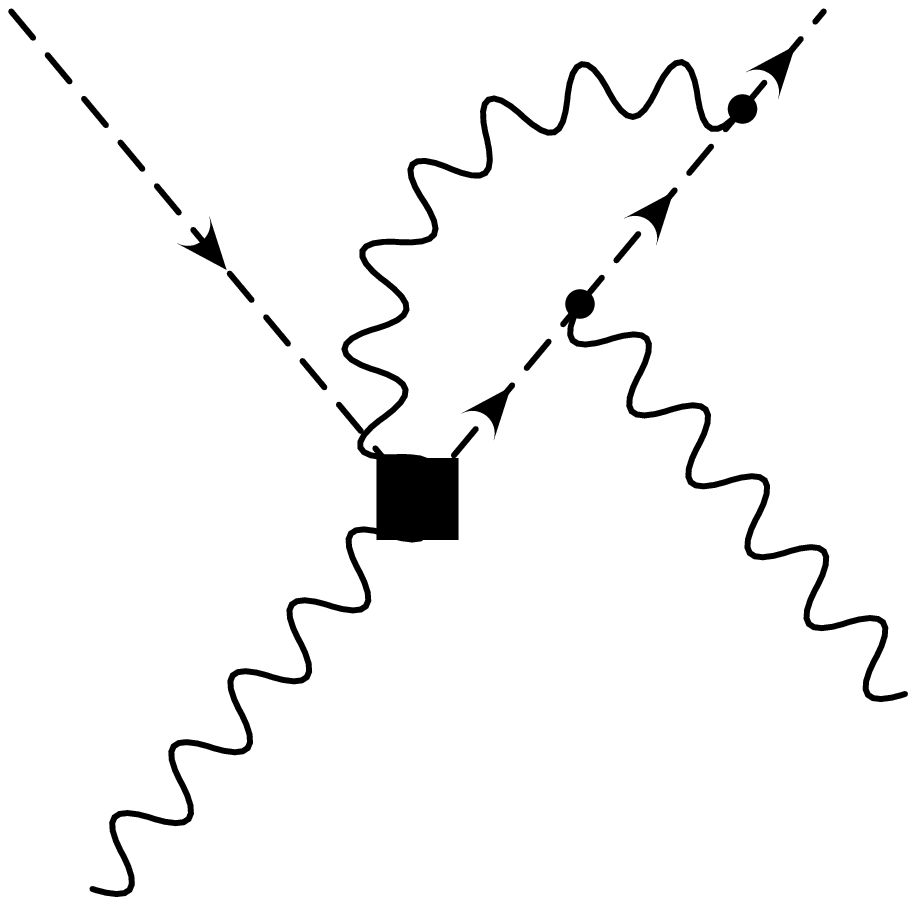} \hspace{0.2cm} & \hspace{0.2cm} \includegraphics[height=3cm]{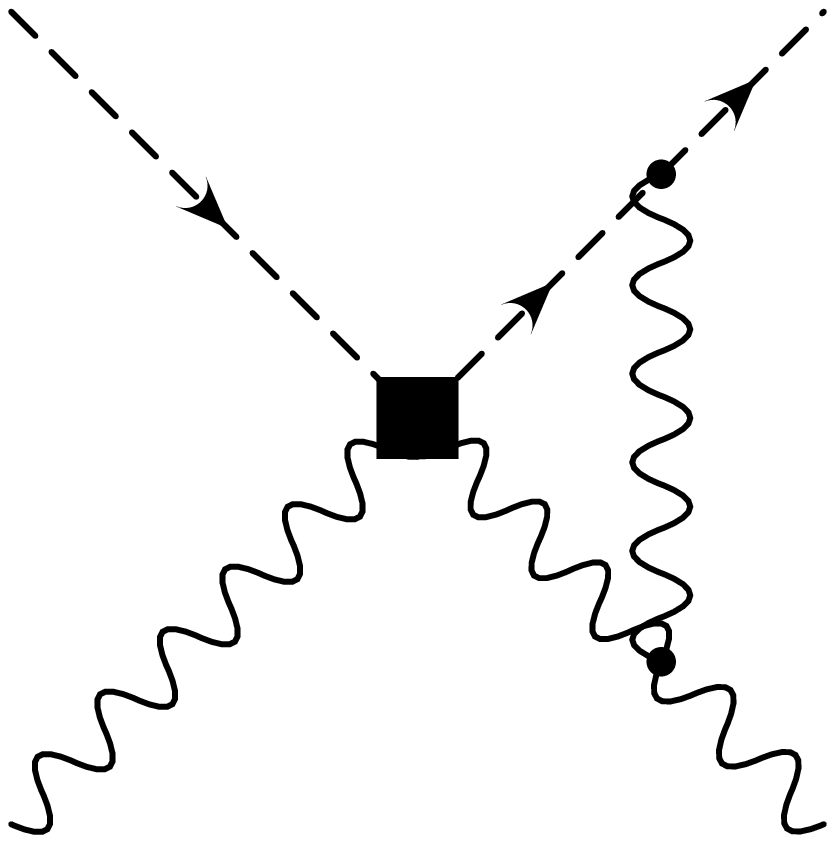} \hspace{0.2cm} &
\hspace{0.2cm} \includegraphics[height=3cm]{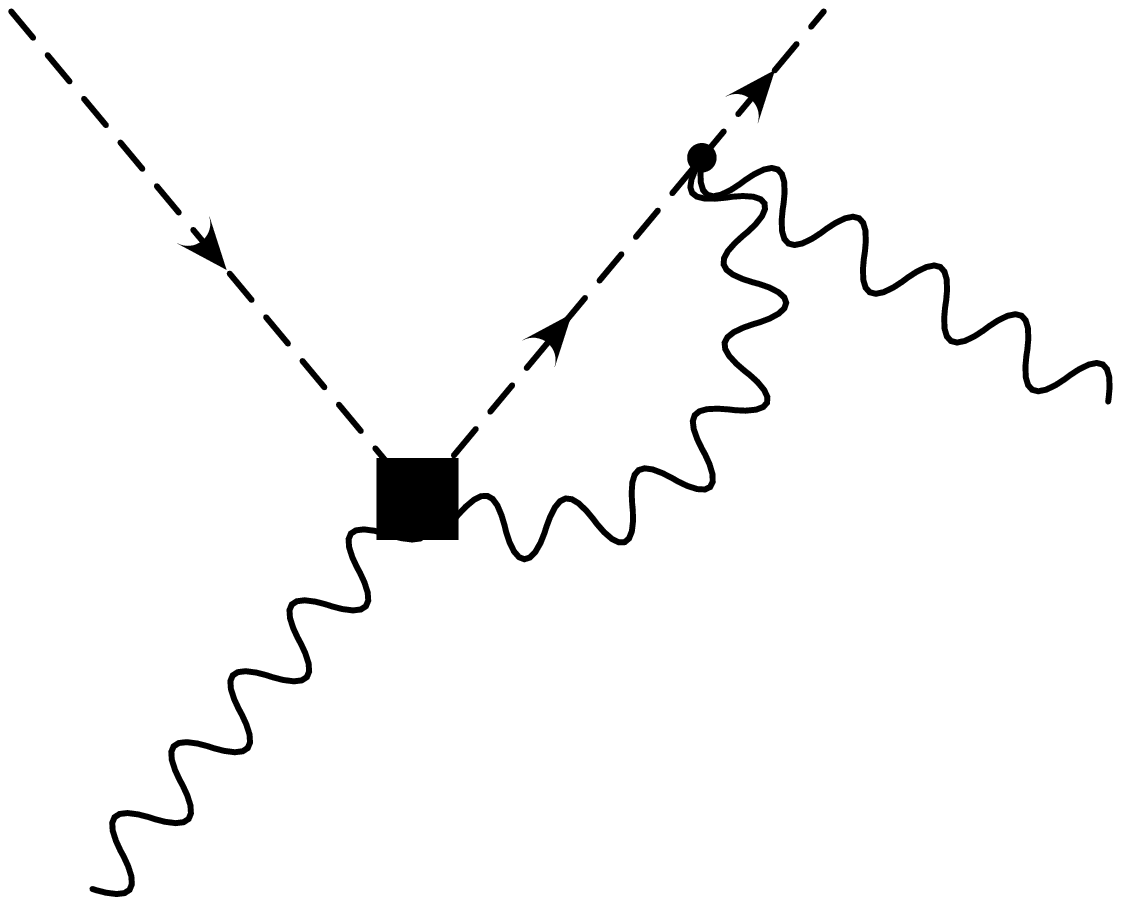} \hspace{0.2cm} & \hspace{0.2cm}  \includegraphics[height=3cm]{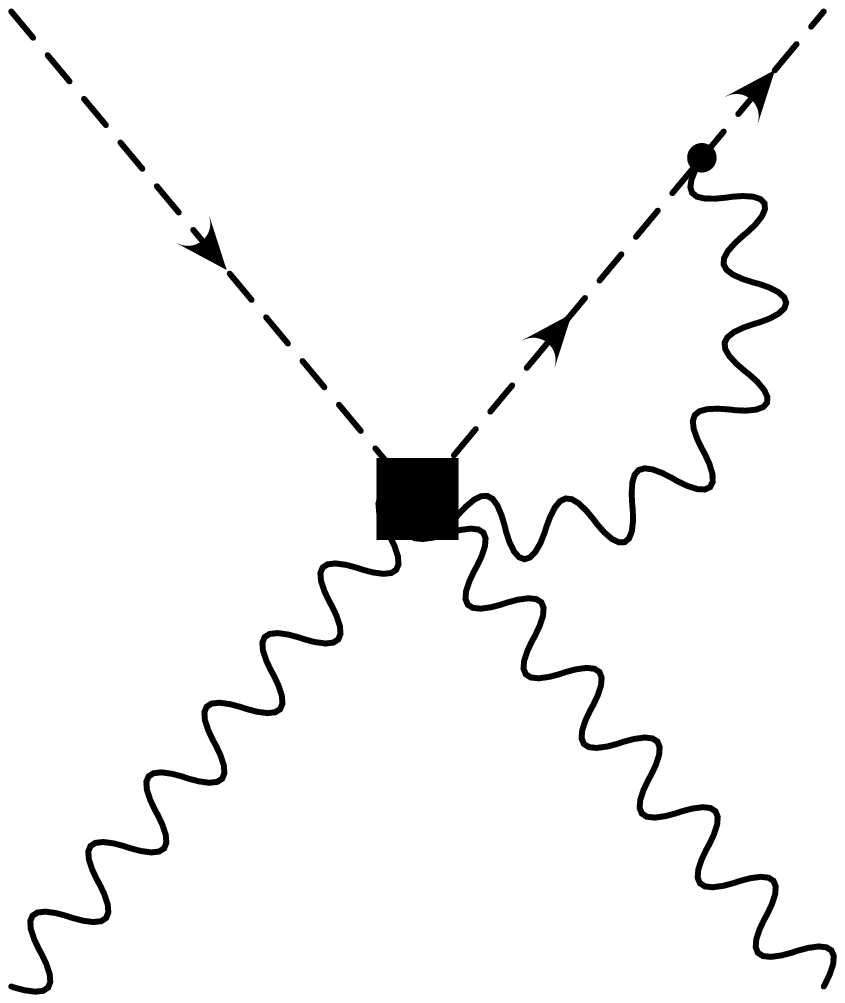} \hspace{0.2cm} \\
(g) & (h) & (i) & (j) \\
\end{array}
\end{eqnarray*}
\end{center}
\caption{\label{fig:graphs} One-loop diagrams for the  renormalization of the operators in Eq.~(\ref{ops}). Graph (e) has a partner graph where the loop is on the other gauge boson line. Graphs (g,h,i,j) have partner graphs where the gauge bosons couple to the incoming scalar line.  Wavefunction graphs have not been shown. Here, the complex scalar field is 
shown as a dashed line, while the gauge fields are shown as wavy lines; in each diagram, the gauge fields are the $B$ or $W^a$ fields depending on the operator considered.}
\end{figure}

The one-loop graphs contributing to the anomalous dimension matrix are shown in Figure~\ref{fig:graphs}.
(We have not shown the ghost graph that vanishes when using dimensional regularization.)
The graphs can be divided into three groups: the graphs of the first row $(a,b)$ couple only to the Higgs field part of the $d=6$ operator insertion; the graphs of the second row $(c,d,e,f)$ couple only to the gauge part of the $d=6$ operator; and the graphs of the third row $(g,h,i,j)$ couple to both the Higgs and gauge fields of the $d=6$ operator insertion. The RGE for the $CP$-even operators at one-loop split into two groups
\begin{subequations}\label{rge}
\begin{align}
\mu \frac{\rd}{\rd \mu} c_G &= \gamma_G\ c_G, \\[10pt]
\mu \frac{\rd}{\rd \mu}\left[ \begin{array}{c}  c_{B} \\  c_{W} \\  c_{W\!B}
\end{array}\right] &= \gamma_{WB} \left[ \begin{array}{c}  c_{B} \\  c_{W} \\  c_{W\!B}
\end{array}\right],
\end{align}
\end{subequations}
where the anomalous dimensions are
\begin{subequations}\label{anomdim}
\begin{align}
\gamma_G &=\frac{1}{16\pi^2}\left[-\frac32 g_1^2-\frac92g_2^2+12\lambda+2Y\right], \\[20pt]
\gamma_{WB} &= \frac{1}{16\pi^2}\left[ \begin{array}{ccc}
\frac12 g_1^2  - \frac{9}2 g_2^2+12\lambda +2Y & 0  & 3 g_2^2  \\[5pt]
0 & -\frac32 g_1^2 -\frac52 g_2^2+12\lambda + 2Y &  g_1^2   \\[5pt]
 2 g_1^2  & 2 g_2^2  & -\frac12 g_1^2 + \frac{9}{2}g_2^2+4\lambda  +2Y
\end{array}\right],
\end{align}
\end{subequations}
and
\begin{align}
Y &= \text{Tr}\left[3 Y_u^\dagger Y_u + 3 Y_d^\dagger Y_d + Y_e^\dagger Y_e\right] \approx 3 y_t^2.
\end{align}
Here, we expand $Y$ in terms of the quark and lepton Yukawa coupling matrices.
Numerically, the top quark Yukawa is the most important contribution to the running. 
The Yukawa coupling correction is a universal correction of the higher dimensional operators due to Higgs wave function renormalization.
The one-loop $\beta$ functions for the coupling constants are given in Appendix~\ref{app:beta}.

The one-loop QCD running of $\mathcal{O}_{G}$ vanishes because a factor of $g_3^2$ is included in the definition of $\mathcal{O}_{G}$.
The sum of graphs $(c,d,e,f)$ vanish for $\mathcal{O}_{W}$ and $\mathcal{O}_{B}$  because $g^2 F^A_{\mu \nu} F^{A\,\mu \nu}$ is not renormalized. This is also trivially
true for the Abelian case, $\mathcal{O}_{B}$, since the graphs don't exist.
The sum does not vanish for $\mathcal{O}_{WB}$, however, since the gauge field part  $g_1 g_2 W^a_{\mu \nu} B^{\mu \nu}$ of 
$\mathcal{O}_{WB}$ is not constrained by a non-renormalization theorem.   The gauge field part of 
$\mathcal{O}_{WB}$ also is not a gauge invariant operator, so the subset of graphs $(c,d,e,f)$  is not gauge invariant for 
$\mathcal{O}_{WB}$.

The renormalization group equations for the $CP$-odd operators are
\begin{subequations}\label{rge2}
\begin{align}
\mu \frac{\rd}{\rd \mu} \widetilde c_G &= \gamma_G\ \widetilde c_G, \\[10pt]
\mu \frac{\rd}{\rd \mu}\left[ \begin{array}{c}  \widetilde c_{B} \\  \widetilde c_{W} \\  \widetilde c_{W\!B}
\end{array}\right] &= \gamma_{WB} \left[ \begin{array}{c}  \widetilde c_{B} \\  \widetilde c_{W} \\  \widetilde c_{W\!B}
\end{array}\right].
\end{align}
\end{subequations}
with the {\it same} anomalous dimensions $\gamma_G,\gamma_{WB}$ as in the $CP$-even case.  The equality of the one-loop $CP$-even and $CP$-odd anomalous dimensions can be understood by the following argument. The $CP$-odd operators involve the product of a field-strength tensor and a dual tensor,
$F_{1\, \mu \nu} \widetilde F^{2\,\mu\nu}$, where the dual can be applied to either $F_1$ or $F_2$. Thus, in computing the graphs, we can choose to apply the dual to the external gauge field that does not participate in the loop, and the graph becomes the same as the $CP$-even case. Because of the freedom of applying the dual to either field-strength, the only graphs where the argument fails are graphs $(c,d,e,f)$ where the loop involves gauge fields from both field-strength tensors.  Now, consider the renormalization of $\widetilde{\mathcal{O}}_B$ and$\widetilde{\mathcal{O}}_{W}$.  The non-renormalization of $g^2 F \widetilde F$ means that the sum of graphs $(c,d,e,f)$ vanishes. For $\widetilde{\mathcal{O}}_B$ these diagrams again trivially do not exist. For the remaining graphs, at most one gauge field takes part in the loop, so the field-strength tensor not including this field can be chosen to be the dual one, and the graph has the same value as the $CP$-even case. For  $\widetilde{\mathcal{O}}_{WB}$, the argument still holds for graphs $(a,b)$ and $(g,h,i,j)$, but there is no non-renormalization theorem for $g_1 g_2 W^a_{\mu \nu} \widetilde B^{\mu \nu}$ to argue that graphs $(c,d,e,f)$ sum to zero. However, since $W^a$ and $B$ gauge fields do not interact with each other, and $B_{\mu \nu}$ is linear in $B_\mu$, graphs $(c,d,f)$ do not exist for $\widetilde{\mathcal{O}}_{WB}$.
Graph $(e)$ must have the two gauge bosons in the loop be $W$ fields from the field strength $W^a_{\mu \nu}$, and the dual can be applied to $B_{\mu \nu}$.  Thus, graph $(e)$ has the same value as the $CP$-even case. This concludes the proof.\footnote{The equality of the $CP$-even and $CP$-odd one loop anomalous dimensions also has been checked by explicit computation.} Clearly, the argument depends crucially on the one-loop structure of the graphs, and will not hold at higher loops. Even the operators $g^2 F^2$ and $g^2 F \widetilde F$ in a non-abelian gauge theory have different anomalous dimensions at two-loop order.

\begin{figure}
\begin{center}
\includegraphics[width=8cm]{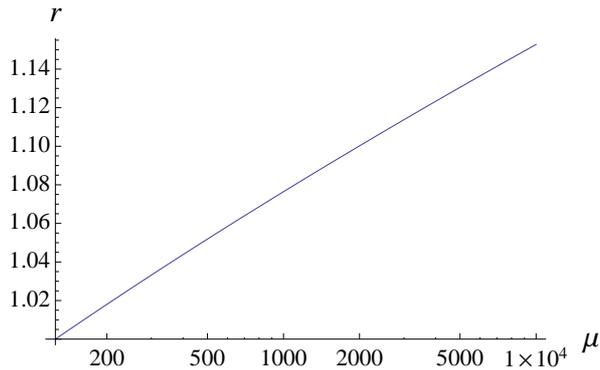}
\end{center}
\caption{\label{fig:rplot} Plot of the top-Yukawa renormalization factor $r(\mu)$ vs $\mu$ in GeV.}
\end{figure}

The renormalization group equations Eq.~(\ref{rge}) need to be integrated between a low-energy scale $\mu$ of order the Higgs mass, and the high-energy scale $\Lambda$ of new physics. The largest contribution to the anomalous dimension is the top quark Yukawa term in Eq.~(\ref{anomdim}), which is proportional to the unit matrix. 
This largest contribution to the anomalous dimension is universal and can be integrated exactly by defining a function $r(\mu)$ which satisfies
\begin{align}
\mu \frac{\rd}{\rd \mu} r(\mu) &= \frac{3 y_t^2(\mu)}{8\pi^2} r(\mu)\,.
\end{align}
Only ratios of $r(\mu)$ enter, so the overall scale of $r$ is irrelevant. A plot of $r(\mu)$ normalized so that $r(\mu=125\,\hbox{GeV})=1$ is shown in Fig.~\ref{fig:rplot}.
The correction is about 8\% to the amplitude for $\mu=1$~TeV.

Writing $c_i(\mu) \equiv r(\mu) d_i(\mu)$, one finds that $d_i(\mu)$ satisfies the renormalization group equation Eq.~(\ref{rge}) with anomalous dimension given by Eq.~(\ref{anomdim}) with $Y \to 0$. Solving this equation to first order in $\log \mu$ gives
\begin{align}
c(M_h) &= \frac{r(M_h)}{r(\Lambda)} \left[ 1 -  \gamma_{WB}(Y \to 0) \ \log \frac{\Lambda}{M_h} \right] c(\Lambda).
\label{running}
\end{align}
This equation is accurate to about 3\% for $\Lambda$ less than 10~TeV. The anomalous dimension matrix 
$\gamma_{WB}(Y \to 0) $ can be evaluated at either $\mu=M_h$ or $\Lambda$ to this order. We will evaluate it at $M_h$ for the numerical results. More accurate results can be obtained by integrating the RGE numerically, but we will use Eq.~(\ref{running}) because it makes the subsequent analysis clearer.

\section{Impact on Phenomenological Studies}\label{pheno}

The RGE improvement of these NP effects can be of significant phenomenological importance, as we will show.  In some cases, these previously neglected RGE effects can be the {\it dominant} contribution of NP to various processes
such as $h \rightarrow \gamma  \gamma$. In this paper, we restrict our attention to two applications of current interest:  the RGE improvement of NP contributions to the partial decay widths $\Gamma(h \rightarrow \gamma  \gamma)$ and $\Gamma(h \to  Z   \gamma)$, and the RGE improvement of global EWPD fit constraints on the dimension-six operator basis.

\subsection{LHC Phenomenology}

The Higgs decay rate $\Gamma(h \rightarrow \gamma  \gamma)$ including $\hsix$ is 
\begin{align}
\label{hgamgammod}
\mu_{\gamma\gamma} \equiv {\Gamma(h \to \gamma \gamma) \over \Gamma^{\text{SM}}(h \to \gamma \gamma)} &\simeq \abs{1- {4 \pi^2 v^2 c_{\gamma \gamma}\over \Lambda^2 I^\gamma}}^2+
\abs{{4 \pi^2 v^2 \tilde c_{\gamma \gamma}\over \Lambda^2 I^\gamma}}^2 ,
\end{align}
following the conventions of Ref.~\cite{Manohar:2006gz},
where $\Gamma^{\text{SM}}$ is the SM rate, and
\begin{align}
c_{\gamma \gamma} &=c_{W}+c_{B} -c_{W\!B}, &
\tilde c_{\gamma \gamma} &=\tilde c_{W}+\tilde c_{B} -\tilde c_{W\!B}\,. 
\end{align}
The SM amplitude is given by $I^\gamma$, which is defined in Appendix~\ref{app:integrals}.
For $\Gamma(h \rightarrow Z  \gamma)$ decay, the  ratio to the standard model rate is
\begin{align}
\mu_{\gamma Z} \equiv { \Gamma(h \to \gamma Z) \over \Gamma^{\text{SM}}(h \to \gamma Z)}&\simeq \abs{1- {4 \pi^2 v^2 c_{\gamma Z}\over \Lambda^2 I^Z}}^2+
\abs{{4 \pi^2 v^2 \tilde c_{\gamma Z} \over \Lambda^2 I^Z}}^2 ,
\label{hZgammod}
\end{align}
where
\begin{align}
c_{\gamma Z} &= c_{W} \cot \theta_W -c_{B} \tan \theta_W -c_{W\!B} \cot 2 \theta_W,  \nn
\tilde c_{\gamma Z} &= \tilde c_{W} \cot \theta_W -\tilde c_{B} \tan \theta_W -\tilde c_{W\!B} \cot 2 \theta_W,
\end{align}
 and the SM amplitude $I^Z$ is defined in the Appendix~\ref{app:integrals}. $I^\gamma$ and $I^Z$ are negative, so we see from Eqs.~(\ref{hgamgammod},\ref{hZgammod}) that $h \to \gamma \gamma$ and $h \to \gamma Z$ are enhanced if $c_{\gamma\gamma}$ and $c_{\gamma Z}$ are positive, and suppressed if they are negative.

The renormalization group improved Higgs decay rates can be computed by using the Wilson coefficients $c_{\gamma \gamma},\tilde c_{\gamma \gamma}, c_{\gamma Z},\tilde c_{\gamma Z}$  in Eqs.~(\ref{hgamgammod}, \ref{hZgammod}) at the scale $\mu \sim M_h$. Using the leading log approximation to the running, given in Eq.~(\ref{running}), one finds
\begin{subequations}\label{eqnmixing}
\begin{align}
\frac{r(\Lambda) c_{\gamma \gamma}(M_h)}{r(M_h)} &= \left[1+\frac{3}{32\pi^2}\left(g_1^2+3g_2^2 -8\lambda\right)\log \frac{\Lambda}{M_h} \right]c_{\gamma \gamma}(\Lambda)
+ \frac{1}{8\pi^2}\left(3g_2^2 -4\lambda\right)\log \frac{\Lambda}{M_h}  c_{W\!B}(\Lambda), \label{mix:a} \\
\frac{r(\Lambda) c_{\gamma Z}(M_h)}{r(M_h)} &= \left[1+\frac{1}{32\pi^2}\left(g_1^2+7g_2^2 -24\lambda\right) \log \frac{\Lambda}{M_h} \right]c_{\gamma Z}(\Lambda)\nn
& + \frac{1}{8\pi^2}\left(g_1 g_2 +4 g_2^2 \cot 2 \theta_W -4\lambda\cot2\theta_W  \right)\log \frac{\Lambda}{M_h}  c_{W\!B}(\Lambda) \nn
&-\frac{\b1 g_1^2-\b2 g_2^2}{16\pi^2}\left(c_{\gamma \gamma}(\Lambda) \sin 2 \theta_W + c_{\gamma Z}(\Lambda) \cos 2\theta_W \right)  \log \frac{\Lambda}{M_h}. 
\label{mix:b}
\end{align}
\end{subequations}
The mixing angle $\theta_W$ in Eq.~(\ref{mix:b}) is evaluated at $\mu=M_h$.
The last term in Eq.~(\ref{mix:b}) comes from the running of the mixing angle $\theta_W$ between $\Lambda$ and $M_h$.  Coefficients  $\b1=-1/6-20 n_g/9=-41/6$, $\b2=43/6-4n_g/3=19/6$ are the leading coefficients of the $g_1$ and $g_2$  $\beta$-functions, and $n_g=3$ is the number of generations. The running of the $CP$-violating operator basis is identical at one loop. 

\subsubsection{A new contribution to $\Gamma(h \rightarrow \gamma  \gamma)$}\label{NewNP}

As shown in Eq.~(\ref{eqnmixing}), the Wilson coefficient $c_{\gamma \gamma}$ at $\mu=M_h$ depends not only on $c_{\gamma\gamma}(\Lambda)$, but also on $c_{W\!B}(\Lambda)$ due to operator mixing. There is no exact symmetry that forbids such mixing in the anomalous dimension matrix.
This mixing provides a demonstration of a new mechanism for a modification of $\Gamma(h \rightarrow \gamma  \gamma)$ due to NP that has
not been considered previously, despite the fact that such effects can be as large as effects of NP which have been examined traditionally. 

When the new physics can be characterized by a single scale $M_\rho$ and a coupling $g_\rho$, simple physical arguments lead to an interesting power counting for the Wilson coefficients of our operator basis~\cite{Giudice:2007fh}.  For coefficients 
${\bar c}_i\equiv c_i v^2/\Lambda^2$, we find the power counting
\begin{eqnarray}
& {\bar c}_B, {\bar c}_W, {\bar c}_{W\!B}, {\bar c}_{DB},{\bar  c}_{DW} \sim O \left( \frac{v^2}{M_\rho^2} \right),\\
&{\bar  c}_G, {\bar c}_{\gamma\gamma}={\bar c}_{W}+{\bar c}_{B} - {\bar c}_{W\!B}, {\bar  c}_{\gamma Z}=\frac{{\bar c}_{W}}{ \tan \theta_W} -{\bar c}_{B} \tan \theta_W -\frac{{\bar c}_{W\!B}}{\tan 2 \theta_W} \sim O \left( \frac{g_\rho^2}{16 \pi^2} \frac{v^2}{M_\rho^2}\right),
\end{eqnarray}
where the last row follows from the fact that the Higgs boson cannot decay to $\gamma \gamma$, $Z\gamma$ and $gg$ at tree-level in any theory that satisfies the minimal coupling assumption.  Note that, when a discrete symmetry is present, there can be further suppression of the operators in the first row, as is the case in R parity conserving SUSY scenarios where there is no tree-level contribution to the $S$ parameter.
Also, if the Higgs boson emerges as a pseudo Nambu-Goldstone boson of the new physics sector, the Higgs decays to $\gamma \gamma$ and $gg$ can only be obtained from a loop that involves couplings which break the global shift symmetry of the pseudo Nambu-Goldstone boson.
In that case, we obtain a further suppression of $g_{SM}^2/g_\rho^2$~\cite{Giudice:2007fh}, so
\begin{eqnarray}
 {\bar c}_G, {\bar c}_{\gamma\gamma} \sim O \left(\frac{g_{SM}^2}{g_\rho^2} \frac{g_\rho^2}{16 \pi^2}  \frac{v^2}{M_\rho^2}\right).
\end{eqnarray}
Here, $g_{SM}$ denotes a combination of the SM couplings $g_{1,2},y_i$. 
The simple power counting above demonstrates the importance of the RGE mixing between the operators we are considering:
\begin{equation}
c_{\gamma\gamma} (\mu)  \sim c_{\gamma \gamma} (\Lambda) + \frac{g_{SM}^2}{16 \pi^2} \log \left(\frac{\Lambda}{\mu}\right) c_i (\Lambda),
\end{equation}
and parametrically the ratio of the RGE contribution over the new physics contribution to $c_{\gamma \gamma}$ scales like
$(g_{SM}^2/g_\rho^2) \log (\Lambda/\mu)$ in the general case and is further enhanced to $\log (\Lambda/\mu)$ in models where the Higgs boson is a pseudo Nambu-Goldstone boson. 
Hence, the RGE effect we want to compute can dominate over the new physics contribution at the matching scale. Similar RGE enhancement is present in the mixing between the operators ${\cal O}_{W\!B}$ and $\partial_\mu |H|^2 \partial^\mu |H|^2$ in Ref.~\cite{Barbieri:2007bh} and has been used to derive some bounds on the deviations of the Higgs couplings to massive gauge bosons from electroweak precision data, see for instance Ref.~\cite{Espinosa:2012ir}. Note that in the case of $c_{Z \gamma}$, the RGE effect is sizeable only in the case of weak coupling $g_\rho \lesssim g_{SM}$.

As a concrete example,  consider the possibility that the Higgs is a pseudo Nambu-Goldstone boson
of a NP sector. Using the SILH formalism of Ref.~\cite{Giudice:2007fh}, a direct matching giving
$c_{\gamma \gamma}(\Lambda)$ is suppressed by a common scale $M^2_\rho$, corresponding to the mass scale of a new strong sector. Here, 
$M_\rho \sim g_\rho \, f$, where $f$ is the analog of the pion decay constant $f_\pi$, and
 $g_{\rho}$ is a coupling in the NP sector with $g_{SM} < g_\rho < 4 \, \pi$. One finds the matching
\begin{align}
c_{\gamma \gamma}(\Lambda) \, \frac{v^2}{\Lambda^2} &\simeq \frac{g^2_{SM}}{8 \, \pi^2 } \, \frac{v^2}{M_\rho^2}.
\end{align}

Now consider the matching onto the operator $\mathcal{O}_{WB}$ due to integrating out the strongly-interacting NP sector. 
It is well known that the $\mathcal{O}_{WB}$ operator receives a tree-level contribution from integrating out new heavy vector bosons.
When this occurs, the interesting possibility arises that the mixing of  $c_{\gamma \gamma}$ with $c_{W\!B}$ due to  renormalization group evolution leads to the \emph{dominant} effect of NP on $h \rightarrow \gamma  \gamma$ decay. This possibility is supported by the fact that the latter mixing effect also is enhanced by a large logarithm.
Integrating out such new spin-one resonances that induce this operator at tree level, one expects $S \sim 4\pi v^2/M_{\rho}^2$, which gives
\begin{align}
c_{W\!B}(\Lambda) \, \frac{v^2}{\Lambda^2} &\simeq - \frac{v^2}{2 \, M_\rho^2}.
\end{align}
Typically, $S$ is positive, and $c_{W\!B}$ is negative.
From these matchings, one sees that at the scale $M_h$ (neglecting the correction to $c_{\gamma\gamma}$ in Eq.~(\ref{mix:a})),
\begin{align}\label{PGHmatch}
c_{\gamma \gamma}(M_h) \, \frac{v^2}{\Lambda^2} & \simeq \frac{r(M_h)}{r(\Lambda)} \,\left[ 2 g^2_{SM}-
(3 \,  g_2^2 - 4\lambda)  \, \log \frac{M_\rho}{M_h} \right] \, \frac{v^2}{16 \, \pi^2 \, M_\rho^2}\,.
\end{align}
Couplings $g_{SM}$ and $g_2$ are of comparable size, so $(3/2) \log\left({M_\rho}/{M_h}\right) \gg1$ is the degree to which the new contribution
to $\Gamma(h \rightarrow \gamma  \gamma)$ dominates over the previously known contribution in pseudo Nambu-Goldstone boson Higgs models.
Numerically, one finds that $(3/2) \log\left({M_\rho}/{M_h}\right)  \sim 3$ for $M_{\rho} \sim 1\, {\rm TeV}$, so this term is expected to be the dominant contribution.  Even when the extra suppression $\sim g^2_{SM}$ is not
present, as is the case for non-Goldstone Higgs scenarios, it is reasonable to expect that the RGE driven contribution we have identified will be significant.  Thus, including this effect is of some 
importance in constructing models of NP that attempt to explain any $\Gamma(h \rightarrow \gamma  \gamma)$ deviation.
Obviously, a similar point holds for future studies of $\Gamma(h \rightarrow Z  \gamma)$ as well, as can be seen from Eq.~(\ref{mix:b}). Note that
negative values of $c_{W\!B}$ lead to a suppression of $\mu_{\gamma\gamma}$ and $\mu_{\gamma Z}$.

The matching condition ${S} \sim 4\pi v^2/M_{\rho}^2$ is a rough estimate based on dimensional grounds. More precise matching conditions based on specific assumptions about the unknown spectral function of
the vector resonances can be utilized if desired, and the conclusions are not significantly changed. For recent calculations along these lines, see
Ref.~\cite{Orgogozo:2012ct, Pich:2012dv}, where, in the framework of minimal composite Higgs models (MCHM~\cite{Agashe:2004rs}), more precise matchings are determined. The finite terms determined in Ref.~\cite{Orgogozo:2012ct, Pich:2012dv} for the particular examples considered in these papers
affect this argument with roughly a further loop factor suppression, so they do not
significantly modify the conclusions.  Nevertheless, due to the general requirement of 
assuming a form of the unknown spectral function that dictates the matching onto $c_{W\!B}(\Lambda)$, strong conclusions are not possible
in a model-independent fashion.\footnote{In particular the results of Ref.~\cite{Orgogozo:2012ct} use $\mu \ll M_{\rho}$ which is associated with a flat spectral density
used in the calculation, as opposed to our matching at $\mu = M_\rho$.}  

One should  note that there are also other modifications to $\Gamma(h \rightarrow \gamma  \gamma)$
in pseudo-Goldstone Higgs models. These effects are discussed in Ref.~\cite{Giudice:2007fh}, and we briefly review them
here for completeness. In this class of models, one also expects modifications of standard model Higgs phenomenology due to NP in a strong sector that induces the following operators
added to the effective Lagrangian
\begin{align}
O_H &= \partial^\mu (H^\dagger \, H) \, \partial_\mu  (H^\dagger \, H), & O_y &= H^\dagger \, H \, \bar{\psi}_L H \psi_R + \hbox{h.c.} ,
\end{align}
with coefficients $c_H/(2 f^2)$ and $c_f \, y_f/f^2$, respectively. These effects are suppressed by the scale $f$, not $M_\rho = g_\rho \, f$, and lead to a suppression of $\Gamma(h \rightarrow \gamma \, \gamma)$ given by \cite{Giudice:2007fh}
\begin{eqnarray}
\frac{\Gamma(h \rightarrow \gamma  \gamma)}{\Gamma(h \rightarrow \gamma  \gamma)_{SM}} = 1 - \frac{v^2}{f^2} {\rm Re} \left[\frac{2 c_t + c_H}{1 + I_W/(N_c Q_t^2 I_t)}+ \frac{c_H}{1 + (N_c Q_t^2 I_t)/I_W} \ \right], 
\end{eqnarray}
neglecting terms suppressed by $g_{SM}^2/g_{\rho}^2$. It is known that one can obtain an enhancement of 
$h \rightarrow \gamma \, \gamma$
if these corrections dominate over the SM contribution, or if $c_t$ is negative, removing the need for any new states giving a large direct matching contribution
(or RGE contribution) to obtain a deviation in $\mu_{\gamma \, \gamma}$. However, at the same time, 
negative $c_t$ diminishes $gg \rightarrow h$ unless it is very large.  See Ref.~\cite{Espinosa:2012im} for a related discussion on the consistency of this possibility with global data.
Of course the parameters that lead to these effects are modified by the inclusion of the RGE effects identified in this paper.

\subsubsection{Inferring the NP scale from RGE modified $\Gamma(h \rightarrow \gamma  \gamma)$}

The measured signal strength for $\gamma  \gamma$ decay in terms of the ratio
to the standard model rate is given by ATLAS as~\cite{ATLAS:2012fk}
\begin{align}
\mu_{\gamma  \gamma} &= 1.80 \pm 0.30 {\rm (stat)}\,^{+0.21}_{-0.15}{\rm (syst)} \,^{+0.20}_{-0.14}{\rm (theory)}, 
\end{align}
for $M_h = 126.6  \pm 0.3 {\rm (stat)} \pm 0.7 {\rm (syst)} \, {\rm GeV}$, while CMS reports~\cite{CMS:2012zwa} 
\begin{align}
\mu_{\gamma  \gamma} &= 1.56 \pm 0.43\,,
\end{align}
for $M_h = 125  \, {\rm GeV}$. 
Neglecting the subtle issues of combining the results of different experiments, and the different central values for the masses of the signal strengths reported,
one finds that a naive combination of these results gives $\mu_{\gamma  \gamma} \simeq 1.7 \pm 0.3$.
If  this excess is attributed to the modifications of the $\Gamma(h \rightarrow \gamma  \gamma)$ amplitude due to $c_{\gamma \gamma}$, one finds two solutions for the central value of $c_{\gamma \gamma}$,
\begin{align}
\frac{v^2}{\Lambda^2}\, c_{\gamma \gamma}(M_h)  &\simeq  -0.1,\ 0.01.
\label{4.13}
\end{align}
The second solution is preferred. The first solution is when $c_{\gamma\gamma}$ switches the sign of the standard model $h \to \gamma \gamma$ amplitude, which
might lead to stability issues of the EW vacuum as discussed in Refs.~\cite{ArkaniHamed:2012kq,Reece:2012gi}.

Further adopting the assumption that the new RGE contribution leads to the observed
enhancement in $\mu_{\gamma  \gamma}$, we can identify the NP scale as a function of $c_{W\!B}$, in this case
\begin{eqnarray}
\frac{r(M_h)}{r(\Lambda)}   \,\frac{v^2}{\Lambda^2} \, \log \frac{\Lambda}{M_h} \ c_{W\!B}(\Lambda) \simeq1.3 \pm 0.5,
\label{4.14}
\end{eqnarray}
using the second solution in Eq.~(\ref{4.13}). The quoted error in the above equation corresponds to the $1 \, \sigma$ range in the naive combined signal strength. The value $\mu_{\gamma  \gamma} \simeq 1.7$ implies either a very low value for the NP scale $\Lambda$, or a large value of $c_{W\!B}$. If 
$\Lambda \sim 1$~TeV, Eq.~(\ref{4.14}) gives $c_{W\!B}(\Lambda) \simeq 11 \pm 4$, compared to the dimensional estimate that $c_{W\!B} \simeq 1$. Alternatively, if $c_{WB}=\pm 1$ then the $h \to \gamma \gamma$ rate increases (decreases) by only 5\% for $\Lambda \sim 1$~TeV. 

It is possible that the current value of $\mu_{\gamma\gamma}$ is biased due to an upward statistical fluctuation (considering the discovery of a Higgs-like scalar as a prior). Eventually, a more accurate measurement of the Higgs decay rate will determine how close $\mu_{\gamma\gamma}$ is to unity. The key question is how large an enhancement of the $h \to \gamma\gamma$ rate is allowed by the RGE contribution, given the current constraints on the $S$ parameter from precision EW measurements. The magnitude of this enhancement sets a benchmark for how accurately $\mu_{\gamma\gamma}$ needs to be measured to rule out NP models at the TeV scale. We examine this question in the next section, using our
determined anomalous dimension to improve the constraints on this operator due to EW precision data.

\subsection{Effect on EWPD and global constraints}\label{ewpd}

The global constraints on the NP operators in Eq.~(\ref{ops}) are of increased interest if deviations in any of the Higgs decays $h \rightarrow \gamma  \gamma, WW, ZZ, Z  \gamma$  becomes statistically significant.  Carefully accounting for the scale dependence of the operators, including the effects of mixing and running, allows a more accurate treatment of global constraints. 
A more precise treatment is particularly important when excesses, such as the current deviations in $\mu_{\gamma  \gamma}$, are being considered as possible hints of new physics. The RGE analysis of Section~\ref{sec:anom} improves these constraints.

In recent global studies \cite{Masso:2012eq,Corbett:2012ja}, the tree-level dependence on $c_{W\!B}$ is eliminated because it is strongly constrained by EWPD. The tree-level equations of motion are used to eliminate $\mathcal{O}_{WB}$ from the operator basis in Ref.~\cite{Masso:2012eq}, while in Ref.~\cite{Corbett:2012ja}, the Wilson coefficient of $\mathcal{O}_{WB}$ is set to zero due to its strong constraint from EWPD. Running the operator basis to other scales, the $\mathcal{O}_{WB}$ operator is regenerated due to mixing if set to zero by hand or eliminated using the equations of motion. The resulting Wilson coefficient will  be loop suppressed, but, because of the sensitivity of EWPD to $\mathcal{O}_{WB}$, this effect can still be phenomenologically relevant. Part of the scale dependence of the operators is captured in the standard equations based on Refs.~\cite{Hagiwara:1993ck,Hagiwara:1993qt,Alam:1997nk} used in these studies. Incorporating the running corrections using $\gamma_{WB}$ includes the full effect of operator mixing for our basis, and  allows more accurate constraints to be drawn in future studies of precision Higgs phenomenology. The RGE analysis includes contributions which were previously omitted, despite being of the same order. 

\subsubsection{$S$ Parameter}

The direct contribution of the $\hsix$ operator basis to the EWPD parameters is known, see Refs.~\cite{Hagiwara:1993ck,Hagiwara:1993qt,Alam:1997nk}.  The $S$ parameter is given by  
\begin{align}
S  &= - \frac{8 \, \pi \, v^2}{\Lambda^2}\left(c_{W\!B}(\Lambda) - \frac{1}{8 \, \pi^2} \, \left[g_2^2 \, c_{W}(\Lambda)+ g_1^2 \, c_{B}(\Lambda) \right] \, \log{\frac{\Lambda}{M_h}}\right),
\label{4.16}
\end{align}
where the $\log{\Lambda/M_h}$ terms come about due to the finite part of the one-loop contribution of 
the $\mathcal{O}_{W},\mathcal{O}_{B}$ operators to $S$.  In contrast to Eq.~(\ref{4.16}), note that Eq.~(\ref{running}) gives
\begin{align}
c_{W\!B}(M_h) &=  \frac{r(M_h)}{r(\Lambda)} \,  c_{W\!B}(\Lambda)\, \left[1 + \frac{g_1^2 - 9 \, g_2^2 - 8 \lambda}{32 \, \pi^2} \, \log{\frac{\Lambda}{M_h}}\right], \nn 
&-  \frac{r(M_h)}{r(\Lambda)} \, \frac{1}{8 \, \pi^2}  \, \left[g_2^2 \, c_{W}(\Lambda)+ g_1^2 \, c_{B}(\Lambda) \right] \, \log{\frac{\Lambda}{M_h}},
\label{4.17}
\end{align}
which makes clear that the $\log{\Lambda/M_h}$ terms in Eq.~(\ref{4.16}) arise from using the formula
\begin{align}
S  &= - \frac{8 \, \pi \, v^2}{\Lambda^2} c_{W\!B}(M_h) \,.
\label{4.18}
\end{align}
However, the  result Eq.~(\ref{4.16}) only includes the second row of Eq.~(\ref{4.17}). The correction in the first row, as well as the top-Yukawa contribution, are not included despite being comparable in magnitude. The value of $S$ consistent with operator renormalization for our basis is given by using Eq.~(\ref{4.18}) and RGE evolution, rather than Eq.~(\ref{4.16}). The answer using the approximate RGE integration of Eq.~(\ref{running}) is to use Eq.~(\ref{4.18}) and Eq.~(\ref{4.17}).\footnote{Note that the effect of the higher dimensional operators in Eq.~(\ref{2.4}) on STU are also included in the analysis of Refs.~\cite{Hagiwara:1993ck,Hagiwara:1993qt,Alam:1997nk}. Until a complete 
one-loop renormalization of the entire operator basis is completed, it is appropriate to use Eq.~(\ref{4.18}) and Eq.~(\ref{4.17}), and to add in the remaining contributions due to the other operators on STU determined in Refs.~\cite{Hagiwara:1993ck,Hagiwara:1993qt,Alam:1997nk}. In particular corrections due to $c_H$ can be significant.The coefficients of these other operators are however unknown until an underlying model is specified. In many models, these additional operators are not as important as those in Eq.~(\ref{ops}), and our relations are sufficient to study EWPD in these classes of models without the additional terms.}

A simple estimate of the impact of these improvements  for phenomenological studies is given by the ratio of $S$ with and without the full one-loop mixing for our operator basis,
\begin{align}
\frac{S_{\text{with}}}{S_{\text{without}}} & \sim \frac{r(M_h)}{r(\Lambda)} \, \left(1 + \frac{g_1^2 - 9 \, g_2^2 - 8 \lambda}{32 \, \pi^2} \, \log{\frac{\Lambda}{M_h}} \right).
\end{align}
Numerically, this ratio ranges from $0.93$ at $\Lambda=500 \, {\rm GeV}$ to $0.86$ at $\Lambda=2.5 \, {\rm TeV}$, which is a significant change in the precision electroweak constraint.

The limit on $S$ gives a constraint on the space of coefficients $c_{W\!B},c_{B},c_{W}$ at the scale $\Lambda$, which in turn constrains the value of $\mu_{\gamma\gamma}$. 
By incorporating the RGE effects we have calculated in this paper, limits on $S$ also are directly related to limits on $\mu_{\gamma\gamma}$ due to EWPD.
The more complete relationship between EWPD and  $\mu_{\gamma\gamma}$ is given in Appendix~\ref{app:reln}.
The numerical version of Eq.~(\ref{C2}) is
\begin{align}
\label{4.21}
\mu_{\gamma\gamma}
&= 1 -  0.02 \, S\, \log \frac{\Lambda}{M_h} + {2.7} \left(\frac{1\, \text{TeV}}{\Lambda}\right)^2\biggl\{ 
1 + 0.0035  \log \frac{\Lambda}{M_h}  \biggr\} c_{\gamma\gamma}(\Lambda) \nn
&\simeq 1 -  0.02 \, S\, \log \frac{\Lambda}{M_h}  + {0.02} \left(\frac{1\, \text{TeV}}{\Lambda}\right)^2 \left(16 \pi^2 c_{\gamma\gamma}(\Lambda)\right)
\end{align}
where the second line emphasizes that $16\pi^2 c_{\gamma\gamma}(\Lambda)$ is expected to be order unity, because $c_{\gamma\gamma}(\Lambda)$ contains a one-loop suppression factor. 
Since $\log \Lambda/M_h$ is positive, enhancements of $\mu_{\gamma\gamma}$ are associated with negative values of $S$. A significant enhancement of $\mu_{\gamma \gamma}$ is associated with a large negative $S$ parameter. A value of $\mu_{\gamma\gamma}=1.73$ implies a large negative value of $S \sim -10$ (when $16 \pi^2{c}_{\gamma\gamma}(\Lambda)$ is set to unity), which is strongly excluded experimentally. The Particle Data Group~\cite{Beringer:1900zz} quotes a value of $S = 0.00^{+0.11}_{-0.10}$ as a result of a fit to $S$, $T$, and $U$. Taking into account correlations, this value leads to $S \leq 0.17$ at $95\%$ C.L for positive values of $S$ --- as expected in many models.

A similar analysis for $\mu_{\gamma Z}$ gives the numerical version of Eq.~(\ref{C4})
\begin{align}
\label{4.21Z}
\mu_{\gamma Z}
&= 1 - 0.014 \, S\, \log \frac{\Lambda}{M_h} +{1.56}   \left(\frac{1\, \text{TeV}}{\Lambda}\right)^2 \biggl\{ \left[
1 + 0.0076\log \frac{\Lambda}{M_h}  \right] c_{\gamma Z}(\Lambda) + 0.012 \log \frac{\Lambda}{M_h} c_{\gamma \gamma}(\Lambda) \biggr\} \nn
&\simeq 1 -  0.014 \, S\, \log \frac{\Lambda}{M_h}   +  {0.01}  \left( \frac{1\, \text{TeV}}{\Lambda}\right)^2 \left[ 
\left(16\pi^2 c_{\gamma Z}(\Lambda)\right) + 0.012 \log \frac{\Lambda}{M_h} \left(16 \pi^2 c_{\gamma \gamma}(\Lambda)\right) \right]\,.
\end{align}

\section{Conclusions}\label{concl}
We have renormalized a subset of the dimension-six operators that encode the impact of NP on the Higgs sector of the SM.
Using these results, we have obtained the RGE  results for the effect of NP on the Higgs decay widths $\Gamma(h \rightarrow \gamma  \gamma)$ and $\Gamma(h \rightarrow Z  \gamma)$ and on the $S$ parameter. We have demonstrated that the leading effect of NP on these decays has not always been properly accounted for in previous studies. In addition, we have shown that the relation between EWPD and the running coefficients of the dimension-six operators contributing to these Higgs decays has not been consistently formulated previously. 

The operator mixing mechanism we have identified makes clear that large excesses in $\mu_{\gamma  \gamma}$ are difficult to reconcile with EWPD constraints,
at least for the operators which we have considered.  Nevertheless, the possibility remains that there are additional RGE effects which we have not computed due
to other dimension-six operators (that we have neglected) which arise from tree-level matching of the new physics and which also mix with $c_{\gamma \, \gamma}$ at one loop. 
Such a scenario could possibly lead to a $\mu_{\gamma \, \gamma}$ enhancement due to the RGE while not being directly constrained by EWPD. 
This mechanism remains a possibility, and it is worthy of future study.  

It also is worth emphasizing the generality of the observations of this paper, which indicates the necessity of a reassessment of the standard expectations for the effects of NP  
on many aspects of one-loop SM Higgs phenomenology.  Our results show that a systematic study of renormalization of the dimension-six operator basis is of crucial importance for the future precision (SM+NP) Higgs physics program. 

Finally, our results also illustrate an important point regarding the global analysis of Higgs signal strengths. An analysis of signal strengths that is framed in terms of a single effective Wilson coefficient for each effective Higgs decay is insufficient to characterize underlying NP models in general. We have shown that the dominant effects can be misunderstood if the scale dependence of the operators is neglected. Conversely, the formalism of a systematic EFT treatment allows one to incorporate the RGE effects that have been shown to have some importance in Higgs phenomenology.

\appendix

\section{$\beta$-functions}\label{app:beta}

The one-loop $\beta$-functions for the standard model couplings are
\begin{align}
\mu \frac{\rd}{\rd \mu} g_1  &= \left(\frac16+ \frac{20}{9}n_g\right) \frac{g_1^3}{16\pi^2}, \nn
\mu \frac{\rd}{\rd \mu} g_2  &= -\left(\frac{43}{6}- \frac{4}{3}n_g\right) \frac{g_2^3}{16\pi^2}, \nn
\mu \frac{\rd}{\rd \mu}  g_3  &= -\left(11- \frac{4}{3}n_g\right) \frac{g_3^3}{16\pi^2}, \nn
\mu \frac{\rd}{\rd \mu} \lambda &= \frac{1}{16\pi^2}\left[24\lambda^2-\lambda \left(3g_1^2+9g_2^2-12y_t^2\right)
+\frac38g_1^4+\frac34g_1^2g_2^2+\frac98g_2^4-6y_t^4 \right], \nn
\mu \frac{\rd}{\rd \mu} y_t &= \frac{1}{16\pi^2}\left[\frac92 y_t^2-\frac{17}{12}g_1^2-\frac94g_2^2-8g_3^2\right]y_t ,
\label{a1}
\end{align}
in the approximation where only the top quark Yukawa coupling is retained. Here $n_g=3$ is the number of generations. The full one-loop and two-loop results can be found in Ref.~\cite{Machacek:1983tz,Machacek:1983fi,Machacek:1984zw,Arason:1991ic}. The couplings $g_1$ and $\lambda$ of Ref.~\cite{Arason:1991ic} (denoted by a prime)
are related to the ones used in this paper by
\begin{align}
g_1^\prime &= \sqrt{\frac{5}{3}}\, g_1, &\lambda^\prime &= 2 \, \lambda \,.
\end{align}
The conventions for $g_2,g_3,y_t,v$ are the same as used here. From Eq.~(\ref{a1}), one finds that the running of the weak mixing angle $\tan\theta_W=g_1/g_2$ is
\begin{align}
\mu \frac{\rd}{\rd \mu} \tan\theta_W  &= \frac{\b2 g_2^2-\b1 g_1^2}{16\pi^2}\tan \theta_W\,,
\end{align}
where $\b1=-1/6-20n_g/9$ and $\b2=43/6-4n_g/3$ are the coefficients of the one-loop $\beta$-functions for $g_1$ and $g_2$, respectively.

We have used the known result for the wavefunction renormalization of the scalar field:
\begin{eqnarray}\label{a4}
Z_{H} = 1 + \frac{(3 - \xi)\, (g_1^2+ 3 \, g_2^2)}{64 \, \pi^2 \, \epsilon}  - \frac{Y}{16 \, \pi^2 \, \epsilon},
\end{eqnarray}
where $\xi$ is the gauge parameter of $R_\xi$ gauge.  The gauge dependence of Eq.~(\ref{a4}) cancels the gauge dependence of the diagrams in Figure~\ref{fig:graphs}.

\section{Feynman Parameter Integrals}\label{app:integrals}

The standard model amplitudes depend on the integration over the Feynman parameter integrals defined in Ref.~\cite{Bergstrom:1985hp}
\begin{eqnarray}
I^\gamma &=& I_W^\gamma\left(\frac{M_h^2}{4M_W^2},0\right) +  \sum_{i } N_i Q_i^2 \left(1-{ \alpha_s \over \pi }\right)\, I_f\left(\frac{M_h^2}{4 m_i^2},0\right),
\end{eqnarray}
where
\begin{eqnarray}
I_f(a,b) &=&\int_0^1 {\rm d}x \int_0^{1-x} \hspace{-0.3cm}{\rm d}y\ {1-4x y \over 1 -4(a-b) xy- 4 b y(1-y)-i0^+},  \\
I^\gamma_W(a,b) &=& \int_0^1 {\rm d}x \int_0^{1-x} \hspace{-0.3cm}{\rm d}y\ {-4+6 xy+4 a x y \over 1 -4(a-b) xy- 4 b y(1-y)-i0^+}.
\end{eqnarray}
In the above equation, the sum on $i$ is over all fermions, and $N_i$ is the number of colors, with $N_i=3$ for quarks and $N_i=1$ for leptons. $Q_i$ is the fermion charge. 
NLO QCD corrections have been included. Similarly the decay to $Z \, \gamma$ depends on the integral
\begin{eqnarray}
I^Z &=& I_W^Z\left( \frac{M_h^2}{4M_W^2},\frac{M_Z^2}{4M_W^2}\right) +   \sum_{i} N_i Q_i  g_i \left(1-{ \alpha_s \over \pi }\right) I_f\left(
\frac{M_h^2}{4 m_i^2},\frac{M_Z^2}{4m_i^2}\right),
\end{eqnarray}
where
\begin{eqnarray}
I^Z_W(a,b) &=& {1\over \tan \theta_W} \int_0^1 {\rm d}x \int_0^{1-x} \hspace{-0.3cm}{\rm d}y\ {\left[5 - \tan^2 \theta_W+2 a\left(1 - \tan^2 \theta_W \right)\right]xy-\left( 3 - \tan^2 \theta_W\right)\over 1 -4(a-b) xy- 4 b y(1-y)-i0^+}\,,\nn
\end{eqnarray}
and $g_i=( T_{3i} -2 \sin^2 \theta_W Q_i)/\sin 2\theta_W$.

The top quark is the dominant fermion contribution for both amplitudes and has the opposite sign from the gauge boson contribution.  One finds $I^\gamma \approx -1.64$ and $I^Z \approx -2.84$ for $M_h = 125 \, {\rm GeV}$. The numerical values were computed using the PDG 2012 \cite{Beringer:1900zz} central values for the standard model parameters, $\alpha_s(M_Z)=0.1184$, $\alpha^{-1}_{\text{em}}(M_Z)=127.944$, $\sin^2\theta_W = 0.23116$, $M_Z=91.1876$\,GeV, $M_W=80.385$\,GeV.

\section{Relation between $h \to \gamma \gamma$, $h \to \gamma Z$ and $S$ }\label{app:reln}

Combining Eq.~(\ref{4.17}) and Eq.~(\ref{4.18}) for the $S$ parameters with Eq.~(\ref{hgamgammod}) and Eq.~(\ref{mix:a}), and keeping terms only to first order in $\log \Lambda/M_h$ gives
\begin{align}
\label{C1}
\mu_{\gamma\gamma}
&\simeq \Biggl|
1 +  \frac{(3 \, g_2^2 -4 \lambda)}{16 \, \pi \, I^\gamma}\, S \, \log \frac{\Lambda}{M_h} - \frac{r(M_h)}{r(\Lambda)} \frac{4 \pi^2 v^2 }{\Lambda^2 I^\gamma} \biggl\{ 
1 +\frac{3 }{32 \pi^2}
\left(g_1^2+3g_2^2-8\lambda\right)   \log \frac{\Lambda}{M_h}  \biggr\} c_{\gamma\gamma}(\Lambda) \Biggr|^2\nn
& +  \abs{\frac{r(M_h)}{r(\Lambda)} \frac{4 \pi^2 v^2 }{\Lambda^2 I^\gamma}}^2 \Biggl|
\left[1 + \frac{3 }{32 \pi^2}
\left(g_1^2+3g_2^2-8\lambda\right)   \log \frac{\Lambda}{M_h} \right]  \widetilde c_{\gamma\gamma}(\Lambda)+ \frac{1}{8\pi^2}\left(3g_2^2-4\lambda\right)  \log \frac{\Lambda} {M_h}  \widetilde c_{W\!B}(\Lambda) \Biggr|^2\,.
\end{align}
When the $c_i$ terms are small compared with the standard model contribution, one can expand this expression retaining only terms linear in $c_i$ to obtain
\begin{align}
\label{C2}
\mu_{\gamma\gamma}
&\simeq 
1 +  \frac{(3 \, g_2^2 -4 \lambda)}{8 \, \pi} {\rm Re}\left(\frac{1}{I^\gamma}\right)\, S \, \log \frac{\Lambda}{M_h}\nn
& - \frac{r(M_h)}{r(\Lambda)} \frac{8 \pi^2 v^2 }{\Lambda^2 } {\rm Re}\left(\frac{1}{I^\gamma}\right)\biggl\{ 
1 +\frac{3 }{32 \pi^2}
\left(g_1^2+3g_2^2-8\lambda\right)   \log \frac{\Lambda}{M_h}  \biggr\} c_{\gamma\gamma}(\Lambda),
\end{align}
where terms proportional to the $\widetilde c_i$ have been neglected.
A similar calculation for $\mu_{\gamma Z}$ gives
\begin{align}
\label{C3}
\mu_{\gamma Z}
&\simeq \Biggl| 1 +   \frac{(g_1 g_2 +4 g_2^2 \cot 2 \theta_W -4 \lambda \cot 2 \theta_W)}{16 \, \pi \, I^Z}\, S \, \log \frac{\Lambda}{M_h} \nn
& - \frac{r(M_h)}{r(\Lambda)} \frac{4 \pi^2 v^2 }{\Lambda^2 I^Z} \biggl\{ 
\left[ 1 + \frac{1 }{32 \pi^2}
\left[\left(1-2\b1 \cos 2 \theta_W\right) g_1^2+\left(7+2\b2 \cos 2 \theta_W\right) g_2^2-24\lambda\right]  \log \frac{\Lambda}{M_h}  \right] c_{\gamma Z}(\Lambda) \nn
&- 
\frac{\sin 2 \theta_W}{16 \pi^2}
\left(\b1 g_1^2-\b2 g_2^2\right)   \log \frac{\Lambda}{M_h}  c_{\gamma\gamma}(\Lambda) \biggr\}\Biggr|^2\nn
&+ \abs{ \frac{r(M_h)}{r(\Lambda)} \frac{4 \pi^2 v^2 }{\Lambda^2 I^Z} }^2 \Biggl|   \left[
1+\frac{1 }{32 \pi^2}
\left[\left(1-2\b1 \cos 2 \theta_W\right) g_1^2+\left(7+2\b2 \cos 2 \theta_W\right) g_2^2-24\lambda\right]  \log \frac{\Lambda}{M_h}  \right] \widetilde c_{\gamma Z}(\Lambda) \nn
&+\frac{1}{8\pi^2}\left( g_1 g_2 +4 g_2^2 \cot 2 \theta_W -4 \lambda \cot 2 \theta_W\right) \log \frac{\Lambda}{M_h}  \widetilde c_{W\!B}(\Lambda) \nn
&-
\frac{\sin 2 \theta_W}{16 \pi^2}
\left(\b1 g_1^2-\b2 g_2^2\right)   \log \frac{\Lambda}{M_h}  \widetilde c_{\gamma\gamma}(\Lambda) \Biggr|^2 .
\end{align}
To linear order in $c_i$, neglecting $\widetilde c_i$, gives
\begin{align}
\label{C4}
\mu_{\gamma Z}
&\simeq 1 +   \frac{(g_1 g_2 +4 g_2^2 \cot 2 \theta_W -4 \lambda \cot 2 \theta_W)}{8 \, \pi  } {\rm Re}\left(\frac{1}{I^Z}\right) \, S \, \log \frac{\Lambda}{M_h} \nn
& - \frac{r(M_h)}{r(\Lambda)} \frac{8 \pi^2 v^2 }{\Lambda^2 }  {\rm Re}\left(\frac{1}{I^Z}\right) \times \nn
&\biggl\{ 
\left[ 1 + \frac{1 }{32 \pi^2}
\left[\left(1-2\b1 \cos 2 \theta_W\right) g_1^2+\left(7+2\b2 \cos 2 \theta_W\right) g_2^2-24\lambda\right]  \log \frac{\Lambda}{M_h}  \right] c_{\gamma Z}(\Lambda) \nn
&- \frac{\sin 2 \theta_W}{16 \pi^2}
\left(\b1 g_1^2-\b2 g_2^2\right)   \log \frac{\Lambda}{M_h}  c_{\gamma\gamma}(\Lambda)\biggr\} .
\end{align}

\acknowledgments

We thank R.~Contino, J.R.~Espinosa,  A.~Pomarol, R.~Rattazzi and F.~Riva and  for insightful discussions.
This research has been partly supported by the European Commission under the ERC Advanced Grant 226371 MassTeV and the contract PITN-GA-2009-237920 UNILHC, by the Spanish Ministry MICNN under contract FPA2010-17747, and by the Department of Energy through DOE Grant No. DE-FG02-90ER40546.

\bibliographystyle{JHEP}
\bibliography{higgs}

\end{document}